\documentstyle[pra,aps,twocolumn,graphicx,epsfig]{revtex}

\begin{document}
\wideabs{
\title{Expansion of a Bose-Einstein Condensate in an atomic waveguide}
\author{L. Plaja$^1$ and L. Santos$^2$}
\address{(1) Departamento de F\'\i sica Aplicada, Universidad de
Salamanca,\\ E-37008 Salamanca, Spain}
\address{(2) Institut f\"ur Theoretische Physik, Universit\"at Hannover, 
 D-30167 Hannover,Germany}
\date{\today}
\maketitle
\begin{abstract}
The expansion of a Bose--Einstein condensate in an atomic waveguide is analyzed.
We study different regimes of expansion,  and 
identify a transient regime between one--dimensional and three--dimensional 
dynamics, in which the properties of the condensate and its further expansion can be well explained by 
reducing the transversal dynamics to a two--level system. 
The relevance of this regime in current experiments is discussed.
\end{abstract}

}
\narrowtext

During the last years, the Bose--Einstein Condensation (BEC) \cite{BEC} 
has constituted one of the most active research fields in modern Atomic Physics. 
Several experiments have shown the coherent character of the outcoupled 
bosons from a condensate \cite{Interferences}, i.e. a BEC constitutes a bright coherent 
matter--wave source, offering promising possibilities in the context of matter--wave optics. 
However, contrary to the extremely dilute non--condensed cold atomic clouds traditionally used 
in Atom Optics \cite{AtomOptics}, the atom--atom interaction induces the BEC dynamics to be  
intrinsically nonlinear. In this sense, the BEC optics has also been 
called Nonlinear Atom Optics (NLAO). Remarkable experiments have 
investigated the resemblance between nonlinear optics phenomena and those 
in NLAO, as four--wave mixing \cite{4WM} and dark--solitons \cite{Solitons}. 
The nonlinearity may deeply alter the dynamics of the wavefunction, and the 
straightforward validity of atom optics elements should be questioned. 
As a consequence, BEC optics should not be considered only as a trivial extension of 
atom optics, but an emerging discipline of its own. 

The analysis of low--dimensional BEC has been recently a
subject of active investigation \cite{Ketterle1D,Safonov,Esslinger,Hannoverfluct,Salomon,Burger}. 
In these systems 
one or two spatial directions do not contribute to the dynamics, since the mean--field energy 
is smaller than the 
typical trapping energy in these directions.  
New effects have been predicted in low--dimensional BEC, as 
quasicondensates \cite{Shlyapnikov2D,Shlyapnikov1D} (for a recent experiment, 
see Ref.\ \cite{Hannoverfluct}), and 
Tonks gas \cite{Shlyapnikov1D,Olshanii}.
In addition, the physics of low--dimensional Bose gases is closely related with one of the 
most actual challenges of the modern Atom optics, namely 
the developing of atom optics micro-structures \cite{birkl,Schmiedmayer}. 
In these devices, an integrated element should be designed to solve 
the fundamental problem of {\em wiring}, i.e. to allow for signal propagation 
between different parts of the structure, with negligible leaks. In this direction, current--carrying
structures \cite{Schmiedmayer} or dipole--force confinement in light fields 
\cite{Ketterle1D,birkl,Bongs} constitute the most promising possibilities.
Up to now, the majority of experiments with atomic waveguides have used 
cold non--condensed atomic clouds. However, recent experiments 
in which a BEC have been confined into an atomic waveguide 
\cite{Bongs} open the interesting and timely question of 
how a BEC expands inside of an atomic waveguide. 
It is the aim of this paper, to characterize in detail the loading of a BEC into a 
waveguide by the expansion of an initially 3D trapped BEC \cite{Bongs}.
The BEC properties (both the initial chemical potential and the subsequent 
expansion dynamics) strongly depend on the dimensionality of the guiding, ranging from 
a purely 3D character to a fully 1D one.
We shall show that due to the cylindrical symmetry of the guiding potential, there 
is an intermediate regime, significant for a large number of experimental situations, in which 
the BEC is well described by considering the radial 
degree of freedom as a two--level system, 
significantly departing from the 1D predictions, even when the
excited radial states are scarcely populated. As a consequence, very
large radial frequencies are necessary to neglect observable 3D effects on the dynamics, 
even for a small number of condensed particles.

In the following, we consider the guiding geometry currently used in
experiments  performed at the University of Hannover \cite{Bongs},  although our study 
may be equally extended to similar geometries, as quantum wires \cite{Schmiedmayer}.
Typically, a BEC is initially created in a hybrid trap, produced by the overlapping of a
cylindrical magnetic trap, and a blue--detuned TEM$_{01}$ (doughnut--mode)
laser beam propagating along the axis  of the magnetic trap. The corresponding
potential exerted on the atoms can be written as  $V({\bf r})=V_i({\bf r})=m\omega_r^2
r^2/2 + m\omega_z^2 z^2/2$, where $\omega_r$ and  $\omega_z$  are the radial
and axial trapping frequency respectively. After the creation (at $t=0$), the
magnetic trap is switched off. The subsequent evolution corresponds,
therefore, to a BEC confined radially by the laser field but free to
move in the axial dimension, as in a waveguide. We shall assume that the
radial frequency is kept  unchanged after the releasing of the magnetic trap,
and therefore for $t>0$,  $V({\bf r}) = V_f({\bf r})= m\omega_r^2 r^2/2$. To facilitate the
discussion of our results, we shall take as a {\em reference} case a $^{87}$Rb
BEC of $10^4$ atoms initially located in an hybrid trap of $\omega_r=2
\pi \times 450$Hz  and $\omega_z=2 \pi \times 10$Hz.  This is a typical
situation in the ongoing experiments performed  at the University of Hannover.

For sufficiently low temperature, the BEC dynamics can be accurately described by the corresponding
Gross-Pitaevskii equation (GPE) \cite{quasibec}
\begin{equation}
i \hbar {{\partial} \over {\partial t}} \psi =
\left \{ - { {\hbar^2} \over {2m}} \nabla^2
+ V + g \left | \psi ({\bf r}, t) \right|^2 \right \}  \psi,
\label{eq:gp}
\end{equation}
where the potential term $V({\bf r}, t)$ is defined above, and $g=4\pi\hbar^2aN/m$ is 
the coupling constant, with $a$ the $s$--wave scattering length 
($5.8 nm$  for $^{87}Rb$), and $m$ the atomic mass, while $N$ is the 
number of condensed atoms. With this choice, $\psi ({\bf r}, t)$ is normalized to $1$.

The BEC expansion in the waveguide after releasing the axial trap is a particular case  
of an anisotropic harmonic trap with time-dependent frequencies. 
For sufficiently shallow traps, the BEC can be considered at any time 
to possess a self--similar 3D Thomas--Fermi (TF) wavefunction 
\cite{castin,kagan1}: 
\begin{equation}
\psi ({\bf r}, t)= b_r^{-1} b_z^{-1/2}\chi(r/b_r,z/b_z) 
 e^{-i \mu_l\tau(t)/\hbar}e^{-i\phi(r,z,t)},
\label{eq:kaganpsi}
\end{equation}
where $\phi(r,z,t) = (m/2\hbar) (r^2 \dot b_r/ b_r+ z^2 \dot b_z/b_z)$, 
$\chi(r,z)^2=(\mu_l -V_i({\bf r}))/g$, 
$b_r(t)$ and $b_z(t)$ are scaling factors which evolve in time according to
${\ddot b}_r(t)+\omega_r^2 b_r(t)=\omega_r^2/b_r^3(t) b_z(t)$ and 
${\ddot b}_z(t)=\omega_z^2/b_r^2(t) b_z^2(t)$,  and  $\tau(t)=\int^t dt'/b_r^2(t')b_z(t')$. 
We shall define the {\em loose}-guiding (LG) regime 
the situations in which the 3D TF approximation is valid. 
In this regime, the initial BEC in the hybrid trap is 
characterized by the corresponding chemical potential
\begin{equation}
\mu_l= {{1} \over {2}} \left( 15 N a \hbar^2 m^{1/2} \omega_r^2 \omega_z \right)^{2/5}.
\label{eq:mul}
\end{equation}

The TF approximation, which neglects the kinetic energy 
term, is roughly correct for trapping energies smaller than the
corresponding self-interaction. 
Since the trapping geometry used in the guiding
experiments is strongly cigar-shaped, the TF approximation may 
become invalid in the transversal directions, 
while still being accurate axially.
We shall refer to this situations as {\em tight}-guiding regime. An extreme 
case of this regime occurs when the radial
binding is strong enough to dominate completely the transversal dynamics,
which can therefore be considered as {\em frozen}. In this case,  the dynamics
becomes 1D in the axial direction, and  the wavefunction can be
factorized as $\psi({\bf r},t)= \psi_{1D} (z, t) \Phi_{0}(r)$, where
$\Phi_{0}(r)$ is the ground--state wavefunction of the transversal 
harmonic oscillator. Provided the TF approximation is valid in the axial coordinate, 
the dynamics of the BEC in the 1D guiding limit can be accurately described by 
a self--similar TF solution in the axial direction, 
\begin{equation}
\psi_{1D} = b_z^{-1/2}\chi_{1D}(z/b_z(t)) 
e^{-i (\mu_{1D}\tau(t)/\hbar + \omega_r t)} e^{-i\phi(z,t)}
\label{eq:kaganpsi1D}
\end{equation}
where $\chi_{1D}(z)^2 = (\mu_{1D}- V(z))/g_{1D}$, 
$\phi(z,t)= m\dot b_z z^2/2\hbar b_z$,
${\ddot b}_z(t)= \omega_z^2/b_z^2(t)$, 
$\tau(t)= \int^t dt'/b_z(t')$, and  $V(z)=m\omega_z^2 z^2/2$.
After integrating over the transversal
direction, one can find the  effective nonlinear parameter 
$g_{1D} =g \int r dr d\phi |\Phi_{0}(r)|^4=(m\omega_r/2\pi\hbar)g$. By
properly normalizing, the 1D chemical potential becomes
\begin{equation}
\mu_{1D}={1 \over 2} \left( { 3\over 2} g_{1D} m^{1/2} \omega_z \right)^{2/3}.
\label{eq:mu1d}
\end{equation}

However, in a general tight--guiding situation, the transversal dynamics 
cannot be considered as frozen, even though
the TF condition is violated in the radial direction. 
Since $\hbar \omega_r > g |\psi|^2$, only a few number of radial excited states is occupied. 
In particular, for a cylindrically symmetric trap only excited states 
of zero polar angular momentum are possible, i.e. 
only one excited state $\Phi_n$ is possible within the $2n$-th radial energy shell.
This produces that for a large number of cases of tight--guiding waveguides (even for 
$\mu \sim 2\hbar\omega_r$), only the 
first excited state $\Phi_1$ is significantly populated before the opening of the hybrid trap. 
In that case the transversal degree of freedom can be well described by a two--level system 
formed by the first two radial eigenstates:
\begin{equation}
\psi({\bf r},t)= C_0(z,t) \Phi_{0} + C_1(z,t) \Phi_{1}.
\end{equation}
Due to the tight--guiding, $|C_1|/|C_0| \ll 1$. Expanding up to first order in
this small parameter, and assuming TF axial profile, the GPE at
$t=0$ can be cast to the form 
\begin{eqnarray}
\tilde \mu_t C_0 & = & V(z) C_0 + g_{1D} C_0^3 - {{3g_{1D}} \over 2} C_0^2 C_1,\\
\tilde \mu_t C_1 & = & (2 \hbar \omega_r + V(z)) C_1 - {g_{1D}\over 2}
C_0^3 + { {3g_{1D}} \over 2}C_0^2 C_1. 
\end{eqnarray}
$C_0$ and $C_1$ are taken as real functions,
and $\tilde \mu_t=\mu_t - \hbar \omega_r$, with 
 $\mu_t$ is the chemical potential in the tight--guiding regime. 
Keeping only first order terms in
$|C_1|/|C_0|$, we find that the
axial profile $n(z)=\int dxdy |\psi(x,y,z)|^2$  can be written in the form 
\begin{equation}
n(z)\simeq n_{1D}(z)\left [ 1+{3 \over 4} \left ( {{n_{1D}(z)} \over
{2\hbar\omega_r/g_{1D}-n_{1D}(z)} }   \right ) \right ],
\end{equation}
where $n_{1D}(z)=|\chi_{1D}(z)|^2$ is the the axial density
in the 1D limit, while the second term
within the brackets introduces the correction coming from the transversal
dynamics. The value of the chemical potential can be found by imposing normalization to the wavefunction:
\begin{eqnarray}
1  &=& \int n(z) dz = {2^{5/2} \over 3} {{\tilde \mu^{3/2}} \over {g_{1D} m^{1/2} \omega_z}} \times 
\nonumber \\
& & \left\{ 1 + { 9
\over 16}  \left[ { 2 \over {\sqrt{\beta}}} (\beta+1)^2 \arctan \sqrt{{ 1 \over \beta}} - 2 \beta - {
{10} \over 3} \right] \right\},
\label{eq:mut}
\end{eqnarray}
where $\beta=2\hbar\omega_r/\mu_{1D}-1$. 
The root $\tilde \mu$ of this equation can be found easily by graphical methods. 
For $\tilde \mu_t/2\hbar\omega_r \ll 1$ one can find the analytic correction to the 1D solution:
\begin{equation}
\tilde \mu \simeq \mu_{1D} \left (
1-\mu_{1D}/5\hbar\omega_r  \right ).
\label{eq:mut2}
\end{equation}

\begin{figure}[ht]  
\begin{center}\  
\epsfxsize=6.2cm  
\hspace{0mm}  
\psfig{file=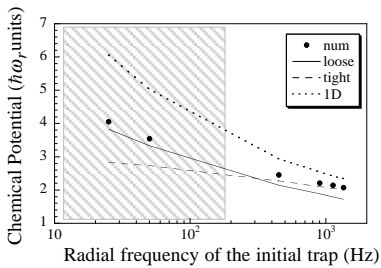,width=6.2cm}\\[0.1cm]  
\end{center}  
\caption{Chemical potential as a function of the radial frequency, $\omega_r/ 2 \pi$. The LG region is shaded.}  
\label{fig:1}   
\end{figure}

In order to test the above discussion, we have numerically solved the 3D GPE 
by means of a Crank--Nicholson method. After obtaining the ground--state of the hybrid trap
using imaginary time evolution, we have calculated the corresponding chemical potential $\mu$
and compared it with the analytical predictions for the different
guiding regimes (Eqs. (\ref{eq:mul}), (\ref{eq:mu1d}) and (\ref{eq:mut})),  
for different trapping parameters and number of condensed atoms. Fig.\ \ref{fig:1} shows the 
dependence of $\mu$ with respect to the 
radial trap frequencies, for the reference experiment discussed above.
As expected, (\ref{eq:mul}) becomes inaccurate in the
tight--guiding regime, and vice versa (\ref{eq:mu1d}) provides wrong results 
for the LG region (shaded region). 
It becomes clear from the figures that the
corrected chemical potential, which accounts for the two--level transversal
dynamics,  describes very well the results obtained by direct numerical
integration of the 3D GPE out of the LG region. In particular, it
accounts very well for the transition region between the 3D and the 1D regime,
and it will merge  with the 1D result for low chemical potentials. It is
particularly interesting that a radial  compression does
not easily lead to a fully 1D behavior (even for $10^4$ atoms),
unless  very large compressions ($\omega_r>2\pi\times 10^4$ Hz) are applied. 
On the contrary, there is a large regime of trap aspect ratios 
($\omega_r=2\pi\times 500$ Hz$\rightarrow 2\pi\times 10^4 $ Hz)
for which Eq. (\ref{eq:mut})
describes the actual chemical potential of the BEC. We should note that
the 1D  prediction can significantly overestimate the chemical potential. As an
example, for $\omega_r=2\pi\times 1.35$kHz, $\mu_{1D}$ is $10\%$ larger than the actual one, even taking into account that more than $95\%$ of the atoms are in the radial
ground state. This difference will have significant consequences in the BEC 
expansion after the releasing of the axial trap, as discussed below.

\begin{figure}[ht]  
\begin{center}\  
\epsfxsize=8.2cm  
\hspace{0mm}  
\psfig{file=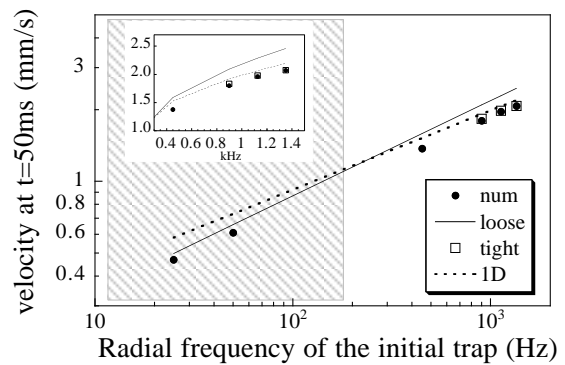,width=8.2cm}\\[0.1cm]  
\end{center}  
\caption{Velocity of expansion after 50ms as a function of the radial frequency, $\omega_r/ 2 \pi$. The LG region is shaded.
Inset: detail of the tight--guiding region. Hollow squares indicate the solution of (\ref{eq:c0dot}) and
(\ref{eq:c1dot}) in the the two--level regime.}  
\label{fig:2}   
\end{figure}

When the magnetic trap is released, the BEC expands axially within the 
waveguide produced by the doughnut laser, until reaching a stationary mean velocity. 
We have numerically simulated this process by solving equation (\ref{eq:gp})  in real time.
We shall concentrate in the dynamics of the BEC variance, defined as 
$\sigma=\sqrt{\langle z^2 \rangle}$. For the LG regime:
$\sigma_{l}(t)=\sigma_{l}(0)b_z(t)/\sqrt{7}$, where 
$\sigma_{l}(0)=a_z (15 N (a/a_z)(\omega_r/\omega_z)^2)^{1/5}$, with $a_z=\sqrt{\hbar/m\omega_z}$, whereas 
for the 1D regime: 
$\sigma_{1D}(t)=\sigma_{1D}(0)b_z(t)/\sqrt{5}$, with 
$\sigma_{1D}(0)=a_z (3 N (a/a_z) (\omega_r/\omega_z))^{1/3}$.
Therefore, at a given time,  
$v=d\sigma_z/dt$ has different laws for the different regimes: 
$v_{l}\sim N^{1/5}\omega_r^{2/5}$ and $v_{1D}\sim N^{1/3}\omega_r^{1/3}$.
Fig.\ \ref{fig:2} shows the variation of $v$ at $50$ ms,
for the cases considered in Fig.\ \ref{fig:1}. 
At this $t$, $v$ has reached an almost stationary value. 
From the figure, it becomes clear the transition between the 3D behavior and the 1D one.  
However, as in Fig.\ \ref{fig:1}, the regime for which the 3D effects on the dynamics become 
unobservable is not easily achievable by increasing $\omega_r$. 
This point becomes clear in the inset of Fig.\ \ref{fig:2}, 
where the numerical results are shown to be parallel to the 1D ones for a large range of radial trap frequencies.
In particular, $v$ can be significantly lower than $v_{1D}$ even for 
large compressions. From the results of the previous section, this effect can be explained by 
(i) the reduction of the initial chemical potential due to the non--negligible influence 
of the population of the first transversal state, and (ii) the departure from the TF profile due to the same reason.
As an example, for $\omega_r=2\pi\times 1.35$kHz, at $50$ ms $v_{1D}-v\simeq 0.15$ mm/s, which determines 
that the condensate variance becomes approximately 
$7\mu$m smaller than that expected from a 1D solution. Such difference is 
experimentally observable with the current imaging techniques.
In the two--level regime, Eq.\ (\ref{eq:gp}) is equivalent to the 
system of time--dependent differential equations:
\begin{eqnarray}
i\hbar\dot C_0 &=& 
\left ( -\frac{\hbar^2}{2m}{{\partial^2} \over {\partial z^2}} + V(z) -\tilde\mu +g_{1D}|C_0|^2 \right ) C_0 \nonumber
\\ &-& g_{1D} |C_0|^2 C_1-\frac{1}{2}g_{1D} C_0^2 C_1^\ast, \label{eq:c0dot} \\
i\hbar\dot C_1 &=& 
\left ( -\frac{\hbar^2}{2m}{{\partial^2} \over {\partial z^2}} + V(z) + 2\hbar\omega_r -\tilde \mu +g_{1D}|C_0|^2
\right ) C_1 
\nonumber \\
&-&\frac{1}{2} g_{1D} |C_0|^2 C_0+\frac{1}{2}g_{1D} C_0^2 C_1^\ast . \label{eq:c1dot}
\end{eqnarray}
The numerical simulation of the previous equations, also depicted in Fig.\ \ref{fig:2}, shows an 
excellent agreement with the direct integration of Eq.\ (\ref{eq:gp}).  
Therefore, the numerically hard task of calculating the 3D corrections by solving the 3D GPE, 
can be performed in a much simpler way, by simply solving a set of $4$ 1D coupled differential 
equations (for $C_0$, $C_0^\ast$, $C_1$ and $C_1^\ast$). 

\begin{figure}[ht]  
\begin{center}\  
\epsfxsize=6.2cm  
\hspace{0mm}  
\psfig{file=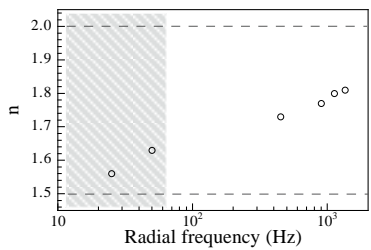,width=6.2cm}\\[0.1cm]  
\end{center}  
\caption{Coefficient $n$ (see text) as a function of the radial frequency, $\omega_r/ 2 \pi$. The LG region is shaded.}  
\label{fig:3}   
\end{figure}

Finally we shall analyze the acceleration of the expansion. It is easy to find that for the 1D regime
\begin{equation}
\ddot \sigma_{1D}= { 3 \over 5^{3/2}} {{\hbar \omega_r a N} \over m} { 1 \over \sigma_{1D}^2}.
\label{eq:acc-sigma1D}
\end{equation}
On the other hand, since
$\omega_r>\omega_z$ in a general guiding situation, we may consider $b_r$ to follow adiabatically
$b_z$, and in the LG regime $b_r(t)=b_z(t)^{-1/4}$. Then,
\begin{equation}
\ddot \sigma_{l}= { 1 \over 7^{5/4}} {{(15 \hbar^2 \omega_r^2 a N)^{1/2}} \over m} { 1 \over
\sigma_{l}^{3/2}}.
\label{eq:acc-sigmal}
\end{equation}
Therefore, in general, the acceleration decreases as $\ddot \sigma(t) \sim 1/\sigma^n(t)$, 
with $n$ ranging from $n=3/2$  in the LG condition and $n=2$ in the 1D regime.
Fig.\ \ref{fig:3} shows $n$ obtained from the 3D GPE, for the cases 
of Fig.\ \ref{fig:1}.

In the present paper, we have analyzed the chemical potential, and 
characterized the expansion dynamics of a BEC within a waveguide. 
We have numerically investigated the transition between a 3D situation (loose guiding) and a 1D one. 
Although the extreme 3D and 1D situations are well described by the
corresponding TF self--similar solutions, there is a broad region
of intermediate parameters which is not well accounted by these solutions.
Due to the cylindrical symmetry of the guiding potential, we have shown that in this region the transversal
dynamics of the BEC can be  modeled as a two--level system, formed by the ground and first
excited state of the radial trapping potential. This model is in 
good agreement with the direct numerical simulation of the corresponding
3D GPE. As we demonstrate, even a small 
population in the first radial excited state can introduce observable differences
between the actual situation and the 1D one,
in the resulting chemical potential, and in the subsequent expansion of the
BEC after releasing of the axial trap.  
From our results, we can 
predict that a very tight transverse confinement is needed to neglect observable 
3D effects determined by the transversal two--level system. In particular, 
this two--level structure could have important effects in the 
reflection and beam--splitting of BEC in waveguides, due to the 
enhancement of the nonlinearity at the edge of the optical elements, 
which could re--couple the transversal levels.  
The properties of a BEC under such conditions will be the subject of further investigations.

We acknowledge  support from  the DFG (SFB
407), the EU TMR Network 'Coherent Matter Wave Interactions', 
the PESC BEC 2000+ , the ESF-FEMTO program, the Spanish Ministerio de Educaci\'on y Cultura (grant EX98-35084508), 
the Spanish Direcci\'on General de Ense\~nanza Superior e Investigaci\'on Cient{\'\i}fica  (PB98-0268), 
and the Consejer{\'\i}a de Educaci\'on y Cultura of the Junta de Castilla y Le\'on
(Fondo Social Europeo, grants SA58/00B, SA044/01 and SA080/02). L. S. wishes to thank the Alexander von Humboldt
Foundation (under the Sofja Kovalewskaja Programm). We thank J. J. Arlt, D. Hellweg, S. Dettmer, M. Lewenstein, A.
Sanpera, and K. Sengstock for useful discussions. L. S. thanks the Optics Group of the University of Salamanca for
hospitality.

\end{document}